\documentclass[twoside]{ilcws07}
\usepackage[latin1]{inputenc}
\usepackage[dvips]{graphicx,epsfig,color}
\usepackage{wrapfig,rotating}
\usepackage{amssymb,amsmath,array}
\begin{document}
\title{LHC/ILC interplay for challenging SUSY scenarios}

\author{K. Desch$^1$, J. Kalinowski$^2$,
G. Moortgat-Pick$^{3}$\footnote{g.a.moortgat-pick@durham.ac.uk,
speaker \cite{url}.}, K. Rolbiecki$^{2}$, W.J. Stirling$^{3}$
\vspace{.3cm}\\
1- Physikalisches Institut, Universit\"at Bonn\\ D-53115 Bonn, Germany
\vspace{.1cm}\\
2- Instytut Fizyki Teoretycznej, Uniwersytet Warszawski\\ PL-00681
Warsaw, Poland
\vspace{.1cm}\\
3- IPPP, Institute for Particle Physics Phenomenology,
University of Durham\\
South Road, Durham DH1 3LE, UK
}

\maketitle
\begin{abstract}
Combined analyses at the Large Hadron Collider and at the
International Linear Collider are important to unravel a difficult
region of supersymmetry that is characterized by scalar SUSY
particles with masses around 2~TeV. Precision measurements of
masses, cross sections and forward-backward asymmetries allow to
determine the fundamental supersymmetric parameters even if only a
small part of the spectrum is accessible. Mass constraints for the
heavy particles can be derived.
\end{abstract}

\section{Introduction}
\label{intro} Supersymmetry (SUSY) is one of the best-motivated
candidates for physics beyond the Standard Model (SM). If
experiments at future accelerators, the Large Hadron Collider (LHC)
and the International Linear Collider (ILC), discover SUSY they will
also have to determine precisely the underlying SUSY-breaking
scenario. Scenarios where the squark and slepton masses are very
heavy (multi-TeV range) as required, for instance, in focus-point
scenarios (FP)~\cite{focuspoint}, are particularly challenging. It
is therefore of particular interest to verify whether the interplay
of an LHC/ILC analysis~\cite{Weiglein:2004hn} could unravel such
models with very heavy sfermions. Here we combine only results from
the LHC with results from the $1^{\rm st}$ stage of the ILC with
$\sqrt{s}\le 500$~GeV.

Methods to derive the SUSY parameters at collider experiments have
been worked out, for instance
in~\cite{Tsukamoto:1993gt,Bechtle:2004pc}. In
\cite{Choi:1998ut,Choi:2001ww,Desch:2003vw} the chargino and
neutralino sectors have been exploited at the ILC to determine the
MSSM parameters. However, in most cases only the production
processes have been studied. Furthermore, it has been assumed that
the masses of the virtual scalar particles are already known.
Exploiting spin effects in  the whole production-and-decay process
in the chargino/neutralino sector~\cite{Moortgat-Pick:1998sk}, it
has been shown in~\cite{Moortgat-Pick:1999ck} that, once the
chargino parameters are known, useful indirect bounds for the mass
of the heavy virtual particles could be derived from
forward--backward asymmetries of the final lepton $A_{\rm
FB}(\ell)$.

Here a FP-inspired scenario is discussed that is characterized by a
$\sim$ 2 TeV scalar particles sector~\cite{Desch:2006xp}.
The analysis is performed entirely at the EW scale, without any
reference to the underlying SUSY-breaking mechanism.

\section{Case study at LHC and ILC}
\label{chap2} We study chargino production $e^{-}+e^{+} \to
\tilde{\chi}^{+}_1+\tilde{\chi}^{-}_1$ with subsequent leptonic
$\tilde{\chi}^{\pm}_1 \to \tilde{\chi}^0_1+\ell^{\pm}+\nu$ and
hadronic decays $\tilde{\chi}^{\pm}_1 \to
\tilde{\chi}^0_1+\bar{q}_d+q_u$, where $\ell=e,\mu$, $q_u=u,c$,
$q_d=d,s$. The production process contains contributions from
$\gamma$- and $Z^0$-exchange in the $s$-channel and from
$\tilde{\nu}$-exchange in the $t$-channel. The decay processes are
mediated by $W^{\pm}$, $\tilde{\ell}_{\rm L}$, $\tilde{\nu}$ or  by
$\tilde{q}_{d{\rm L}}$, $\tilde{q}_{u {\rm L}}$ exchange. The masses
and eigenstates of the neutralinos and charginos are determined by
the fundamental SUSY parameters: the $U(1)$, $SU(2)$ gaugino mass
parameters $M_1$, $M_2$, the Higgs mass parameter $\mu$ and the
ratio of the vacuum expectation values of the two neutral Higgs
fields, $\tan\beta=\frac{v_2}{v_1}$. In our case study the MSSM
parameters at the EW scale are given by: $M_1=60~\mbox{\rm
GeV},~M_2=121~\mbox{\rm GeV},~M_3=322~\mbox{\rm
GeV},~\mu=540~\mbox{\rm GeV},~\tan\beta=20.$ The derived masses of
the SUSY particles are listed in Table~\ref{tab:1}.
\begin{table}\center
\begin{tabular}{cccccc}
\hline\noalign{\smallskip}
 $m_{\tilde{\chi}^{\pm}_{1,2}}$
& $m_{\tilde{\chi}^{0}_{1,2,3,4}}$ &
 $m_{\tilde{\nu}_e,\tilde{e}_{\rm {R,L}}}$ &
$m_{\tilde{q}_{\rm {R,L}}}$ & $m_{\tilde{t}_{1,2}}$ &
$m_{\tilde{g}}$ \\
117, 552 & 59, 117, 545, 550  & 1994, 1996, 1998 &  2002, 2008  &  1093, 1584 & 416\\
\noalign{\smallskip}\hline
\end{tabular}
\caption{ Masses of the SUSY particles [in GeV]. \label{tab:1} }
\end{table}

\subsection{Expectations at the LHC}\label{lhc}
All squarks in this scenario are kinematically accessible at the
LHC. The largest squark production cross  section is for
$\tilde{t}_{1,2}$. However, with stops decaying mainly to
$\tilde{g}t$ (with $BR(\tilde{t}_{1,2}\to \tilde{g} t)\sim 66\%$),
where background from top production will be large, no new
interesting channels are open in their decays.

Since the gluino is rather light in this scenario, several gluino
decay channels can be exploited. The largest branching ratio for the
gluino decay in our scenario is a three-body decay into neutralinos,
$BR(\tilde{g}\to \tilde{\chi}^0_2 b \bar{b})\sim 14\%$, followed by
a subsequent three-body leptonic neutralino decay
$BR(\tilde{\chi}^0_2\to \tilde{\chi}^0_1 \ell^+ \ell^-)$,
$\ell=e,\mu$ of about 6\%. In this channel the dilepton edge will be
clearly visible~\cite{Weiglein:2004hn}. The mass difference between
the two light neutralino masses can be measured from the dilepton
edge with an uncertainty of about
$\delta(m_{\tilde{\chi}^0_2}-m_{\tilde{\chi}^0_1}) \sim 0.5~\mathrm{
GeV}$~\cite{Kawagoe:2004rz}. The gluino mass can be reconstructed in
a manner similar to the one proposed in~\cite{Gjelsten:2005aw} and a
relative uncertainty of $\sim$2\% can be expected.

\subsection{Expectations at the ILC}\label{ilc}
At the first stage of the ILC, $\sqrt{s}\le 500$~GeV, only light
charginos and neutralinos are kinematically accessible. However, in
this scenario the neutralino sector is characterized by very low
production cross sections, below 1~fb, so that it might not be fully
exploitable~\cite{Desch:2006xp}. Only the chargino pair production
process has high rates and we use $\sqrt{s}=350$ and $500$~GeV. The
chargino mass can be measured in the continuum, with an error of
about $ 0.5$~GeV~\cite{Martyn:1999tc}. This can serve to optimize
the ILC scan at the threshold which, can be used to determine the
light chargino mass very precisely, to about~\cite{Martyn:1999tc}:
\begin{equation}
m_{\tilde{\chi}^{\pm}_1}=117.1 \pm 0.1 ~\mathrm{ GeV}.
\label{eq_massthres}
\end{equation}
The mass of the lightest neutralino $m_{\tilde{\chi}^0_1}$ can be
derived, either from the lepton energy distribution
($BR(\tilde{\chi}^{-}_1\to \tilde{\chi}^0_1 \ell^-
\bar{\nu}_\ell)\sim 11\%$) or from the invariant mass distribution
of the two jets ($BR(\tilde{\chi}^-_1\to \tilde{\chi}^0_1 q_d
\bar{q}_u)\sim 33\%$). We take~\cite{Martyn:1999tc}
\begin{equation}
m_{\tilde{\chi}^{0}_1}=59.2 \pm 0.2~\mathrm{ GeV}.
\label{eq_masslsp}
\end{equation}
Together with the information from the LHC a mass uncertainty for
the second light neutralino of about
\begin{equation}
m_{\tilde{\chi}^{0}_2}=117.1 \pm 0.5~\mathrm{ GeV}
\label{eq_masschi02}
\end{equation}
can be assumed.

\begin{wrapfigure}{r}{0.5\textwidth}
\centerline{\includegraphics[width=.48\textwidth]{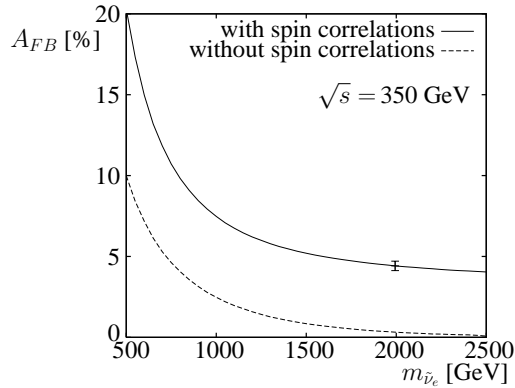}}
\caption{Forward--backward asymmetry of the final
$e^-$~\cite{Desch:2006xp} as a function of $m_{\tilde{\nu}_e}$. For
nominal value of $m_{\tilde{\nu}_e}=1994$ GeV the expected
experimental errors are shown.} \label{fig:1}
\end{wrapfigure}
We identify the chargino pair production process in the fully
leptonic and semileptonic final states and estimate an overall
selection efficiency of 50\%. The $W^+W^-$ production is the
dominant SM background. For the semileptonic (slc) final state, this
background can be efficiently reduced from the reconstruction of the
hadronic invariant mass. In Table~\ref{tab:3}, we list cross
sections multiplied by the branching fraction $B_{slc}=2\times
BR(\tilde{\chi}^+_1 \to \tilde{\chi}^0_1 \bar{q}_d q_u)\times
BR(\tilde{\chi}^-_1 \to \tilde{\chi}^0_1 \ell^- \bar{\nu})+
[BR(\tilde{\chi}^-_1 \to \tilde{\chi}^0_1 \ell^- \bar{\nu})]^2\sim
0.34$ (first two families) including a selection efficiency of
$e_{slc}=50\%$. The error includes the statistical uncertainty of
the cross section and $A_{\rm FB}$ (see~\cite{Desch:2006xp}) based
on ${\cal L}=200$~fb$^{-1}$ in each polarization configuration,
$(P_{e^-},P_{e^+})=(-90\%,+60\%)$ and $(+90\%,-60\%)$, and a
relative uncertainty in the polarization of $\Delta
P_{e^\pm}/P_{e^\pm}=0.5\%$~\cite{Moortgat-Pick:2005cw}.

\section{Parameter determination}
\label{chap3}
We determine the underlying SUSY parameters in several steps:
\subsection{Analysis without {\boldmath $A_{\rm FB}$}}
Only the masses of $\tilde{\chi}^{\pm}_1$, $\tilde{\chi}^0_1$,
$\tilde{\chi}^0_2$ and the chargino pair production cross section,
including the fully leptonic and the semileptonic decays have been
used as observables. A four-parameter fit for the parameters $M_1$,
$M_2$, $\mu$ and $m_{\tilde{\nu}}$ has been applied, for fixed
values of $\tan\beta=5$, 10, 15, 20, 25, 30, 50 and 100. Due to the
strong correlations among parameters~\cite{Desch:2006xp}, fixing of
$\tan\beta$ is necessary. We perform a $\chi^2$ test and obtain the
following $1\sigma$ bounds for the SUSY parameters:
\begin{eqnarray*}
&&59.4  \le M_1 \le 62.2~\mathrm{ GeV},\quad
 118.7  \le M_2 \le 127.5~\mathrm{ GeV},\\
&& 450  \le \mu \le 750~\mathrm{ GeV},\quad 1800 \le
m_{\tilde{\nu}_e}\le 2210~\mathrm{ GeV}.
\end{eqnarray*}
\subsection{Analysis including leptonic {\boldmath $A_{\rm FB}$}}
We now extend the fit by using as an additional observable the
leptonic forward--backward asymmetry, which is sensitive to
$m_{\tilde{\nu}}$. Proper account of spin correlations is crucial,
see Fig.~\ref{fig:1}. The $SU(2)$ relation between the two virtual
masses $m_{\tilde{\nu}}$ and $m_{\tilde{e}_{\rm L}}$ has been
assumed. The multiparameter fit strongly improves the results. No
assumption on $\tan\beta$ has to be made. We find
\begin{eqnarray*}
&&59.7\le M_1\le 60.35~\mathrm{ GeV},\quad 119.9\le M_2 \le
122.0~\mathrm{ GeV},\quad
500\le \mu \le 610~\mathrm{ GeV}, \\
&& 14\le \tan\beta  \le 31,\quad 1900\le m_{\tilde{\nu}_e} \le
2100~\mathrm{ GeV}. \label{eq:leptonic}
\end{eqnarray*}
The constraints for the mass $m_{\tilde{\nu}_e}$ are improved by a
factor of about $2$ and for gaugino mass parameters $M_1$ and $M_2$
by a factor of about $5$. The masses of heavy chargino and
neutralinos are predicted to be within the ranges
\begin{equation*}
506 < m_{\tilde{\chi}^0_3} < 615~\mbox{GeV},\quad
\label{eq_mchi3_pred} 512 < m_{\tilde{\chi}^0_4} < 619
~\mbox{GeV},\quad 514 < m_{\tilde{\chi}^{\pm}_2} < 621~\mbox{GeV}.
\end{equation*}
\subsection{Analysis including hadronic and leptonic {\boldmath
$A_{\rm FB}$}: test of {\boldmath $SU(2)$}}
\begin{table}\center
\renewcommand{\arraystretch}{1.2}
\begin{tabular}{lcccc} \hline
 & \multicolumn{2}{c}{$\sqrt{s}=350$ GeV } &\multicolumn{2}{c}{$\sqrt{s}=500$ GeV }
\\
 $(P_{e^-},P_{e^+})$ & $(-90\%,+60\%)$ & $(+90\%,-60\%)$ &$(-90\%,+60\%)$ & $(+90\%,-60\%)$
\\ \hline
$\sigma(\tilde{\chi}^+_1\tilde{\chi}^-_1)$ &  6195.5 & 85.0 &  3041.5 &  40.3 \\
 $\sigma(\tilde{\chi}^+_1\tilde{\chi}^-_1)\, B_{slc}\, e_{slc} $  & 1062.5$\pm$4.0
& 14.6$\pm 0.7$ & 521.6$\pm 2.3$ &  6.9$\pm 0.4$ \\
 $A_{\rm FB}(\ell^-)$/\% & 4.42$\pm$0.29  & --  & 4.62$\pm$0.41 & -- \\
 $A_{\rm FB}(\bar{c})$/\% & 4.18$\pm$0.74 & --  & 4.48$\pm1.05$ & --  \\ \hline
\end{tabular}
\caption{Cross sections for the process $e^+e^-\to
\tilde{\chi}^+_1\tilde{\chi}^-_1$ [in fb] and forward--backward
asymmetries $A_{\rm FB}$ in the leptonic $\tilde{\chi}^-_1 \to
\tilde{\chi}^0_1 \ell^- \bar{\nu}$ and hadronic $\tilde{\chi}^-_1
\to \tilde{\chi}^0_1 s \bar{c}$ decay modes, for different beam
polarization $P_{e^-},P_{e^+}$. Concerning the errors, see text and
\cite{Desch:2006xp}. \label{tab:3}}
\end{table}
In the last step both the leptonic and hadronic forward--backward
asymmetries have been used. With the constraints for the squark
masses from the LHC, the hadronic forward--backward asymmetry could
be used to control the sneutrino mass. The leptonic
forward--backward asymmetry provides constraints on the selectron
mass and the $SU(2)$ relation between selectron and sneutrino masses
could be tested. A six-parameter fit for the parameters $M_1$,
$M_2$, $\mu$, $m_{\tilde{\nu}}$, $m_{\tilde{e}_{\rm L}}$ and
$\tan\beta$ has been applied, resulting in the following
constraints:
\begin{eqnarray*}
&&59.45\le M_1\le 60.80~\mathrm{ GeV},\quad
 118.6\le M_2\le 124.2~\mathrm{ GeV},\quad
420\le \mu\le 770~\mathrm{ GeV},\\
&& 11\le \tan\beta  \le 60,\quad 1900\le m_{\tilde{\nu}_e} \le
2120~\mathrm{ GeV},\quad 1500~\mathrm{GeV}\le m_{\tilde{e}_{\rm L}}.
\label{eq:hadnosu2}
\end{eqnarray*}
The limits are somewhat weaker comparing to the previous case, but
we get now constraints for one additional parameter: the selectron
mass.

\section{Conclusions and outlook}
\label{chap4}
Scenarios with heavy scalar particles are challenging for
determining the MSSM parameters. A very powerful tool in this kind
of analysis turns out to be the forward--backward asymmetry. This
asymmetry is strongly dependent on the mass of the exchanged heavy
particle. If the $SU(2)$ constraint is applied, the slepton masses
can be determined to a precision of about 5\% for masses around
2~TeV at the ILC running at 500~GeV. In addition powerful
predictions for the heavier charginos/neutralinos can be made.

In future developments it will be crucial to add radiative
corrections which are so far available separately for the
production~\cite{production} and decays~\cite{decay}. Full
simulations of the whole production-and-decay process will be
necessary for precision physics at the ILC.

\section*{Acknowledgements} This work was supported in part by
the Polish Ministry of Science and Higher Education Grant
No.~1~P03B~108~30 and by the European Community Marie-Curie Research
Training Network MRTN-CT-2006-035505 (HEPTools).

\begin{footnotesize}

\end{footnotesize}

\end{document}